# Giant magnetoresistance in nanoscale ferromagnetic heterocontacts

A. N. Useinov [1,2], R. G. Deminov [1], L. R. Tagirov [1,2] and G. Pan [2]

[1] Kazan State University, 420008 Kazan, Russian Federation
[2] CRIST, University of Plymouth, Plymouth, Devon PL4 8AA, United Kingdom

A quasiclassical theory of giant magnetoresistance in nanoscale point contacts between different ferromagnetic metals is developed. The contacts were sorted by three types of mutual positions of the conduction spin-subband bottoms which are shifted one against another by the exchange interaction. A model of linear domain wall has been used to account for the finite contact length. The magnetoresistance is plotted against the size of the nanocontact. In heterocontacts the magnetoresistance effect turned out to be not only negative, as usual, but can be positive as well. Relevance of the results to existing experiments on GMR in point heterocontacts is discussed.

## 1. Introduction

The experimental discovery of ultra-high magnetoresistance (MR) in ferromagnetic nanocontacts has attracted considerable attention due to its potential technological applications for future generation of magnetoresistive sensors [1-8]. Two mechanisms of giant magnetoresistance (GMR) in magnetic nanocontacts were proposed to explain the experimental data: one is the enhancement of the impurity scattering in a domain wall (DW) [2,4], and the other is the scattering of electrons by an energy landscape of DW (domain wall scattering) [9-11]. Both mechanisms essentially exploit sharpness of the domain wall profile shrinking into a narrow (ultimately of the atomic size) constriction [12-15]. The DW scattering theory [11] is general enough to admit spin asymmetry of the bulk impurity scattering (conduction electron mean free path in the spin-subbands of a ferromagnet may differ 5 to 7 times [16]) as well as the spin asymmetry of the interface scattering (contacting ferromagnets can be different - ferromagnetic heterocontacts). The aim of the present paper is to analyse influence of the spin-asymmetry of the interface scattering on magnetoresistance of ferromagnetic point heterocontacts and to search for optimal conditions at which the GMR effect can be maximized.

## 2. Conductance of a ferromagnetic heterocontact

We consider a small-area contact between two single-domain ferromagnetic metals. When the magnetization on both sides of the contact is in parallel (P) alignment there is no domain wall in the constriction, and the electric current flows through the point contact independently in each of the conduction electron spin-subband. At an antiparallel (AP) alignment of the magnetisations a domain wall is created in the constriction [12-15]. Simultaneously, the conduction spin-subband assignment in one of the magnetic domains reverses with respect to the previous one. In the case of ferromagnetic heterocontact, the band structures of the spin-subbands of the ferromagnetic metals do not coincide with either spin-up or spin-down conduction electrons. It is obvious that potential barriers at the interface of the contact (see figure insets below) are different for the P and AP alignments. As a result, scatterings of electrons associated with these potential barriers and magnetisation profiles at the interface are different for the two alignments, which gives rise to magnetoresistance.

The case of ferromagnetic homocontact (a contact made of the same ferromagnetic metal) had been considered in Ref. [11] in the quasiclassical approximation. Using the same approach, we give here general derivation of the conductance of ferromagnetic heterocontact made of different ferromagnetic metals. The model of nanocontact we consider is a circular hole of radius $a$ made in an impenetrable membrane, which divides the space into two halves, each of which is occupied by a single-domain ferromagnetic metal. The $z$-axis of the coordinate system is chosen perpendicular to the membrane plane. Our aim is to calculate the electric current $I^z$ through the hole in response to the voltage drop $V$ applied to the outer leads far away from the contact:

$$I^z(z \to 0) = a \int_0^\infty dk J_1(ka) j^z(0,k). \tag{1}$$

Here the Bessel function $J_1(x)$ comes from integration of the current density $j^z(z=0,\rho)$ over the contact cross-section, $j^z(0,k)$ is the Fourier-transform of the current density $j^z(z=0,\rho)$ over the in-plane coordinate $\rho$. The current density can be expressed via the antisymmetric quasiclassical Green function (GF) $g_a(z,\rho)$ as follows ($c = \hbar = 1$):

$$j^z(z,\rho) = -\frac{ep_F^2}{2\pi} \int_0^{\pi/2} d\Omega_\theta \cos\theta\, g_a(z,\rho). \tag{2}$$

The antisymmetric GF itself is a solution of the system of Boltzmann-type equations [17]:

$$l_{z,\alpha} \frac{\partial g_{a,\alpha}}{\partial z} + \mathbf{l}_{\|,\alpha} \frac{\partial g_{s,\alpha}}{\partial \rho} + g_{s,\alpha} - \langle g_{s,\alpha} \rangle = 0,$$

$$l_{z,\alpha} \frac{\partial g_{s,\alpha}}{\partial z} + \mathbf{l}_{\|,\alpha} \frac{\partial g_{a,\alpha}}{\partial \rho} + g_{a,\alpha} = 0, \tag{3}$$

supplied with the boundary conditions (BC) at interfaces

$$g_{aL,\alpha} = g_{aR,\alpha} = \begin{cases} g_{a,\alpha}, & p_\| < p_{FL}, p_{FR} \\ 0, & \min(p_{FL}, p_{FR}) < p_\| \end{cases}, \tag{4}$$

$$2R_\alpha g_{a,\alpha} = D_\alpha \left( g_{sL,\alpha} - g_{sR,\alpha} \right). \tag{5}$$

In the above equations $g_{s(a)} = 1/2[g_\alpha(n_z,z,\rho) \pm g_\alpha(-n_z,z,\rho)]$ is the single-particle quasiclassical Green function symmetric (antisymmetric) with respect to the projection $n_z = p_{z,\alpha}/p_{F,\alpha}$ of the Fermi momentum $p_{F,\alpha}$ on the axis $z$; $l_{z,\alpha} = l_\alpha \cos\theta$ is the projection of the spin-dependent electron mean free path $l_\alpha$ on the axis $z$, $l_{\|,\alpha}^2 = l_\alpha^2 - l_{z,\alpha}^2$; $\alpha = (\uparrow,\downarrow)$ is the spin index, and $\rho = (x,y)$ is the coordinate in the plane of the contact. The angular brackets in $\langle g_s \rangle$ mean averaging over the solid angle: $\langle g_s \rangle = \oint d\Omega/2\pi\, g_s$, $p_\|$ is the projection of the spin-dependent Fermi momentum $p_{F,\alpha}$ on the plane of the contact, $D_\alpha$ and $R_\alpha = 1 - D_\alpha$ are the angular and spin-dependent, quantum-mechanical transmission and reflection coefficients, respectively. Boundary conditions (4) and (5) obey the specular reflection law:

$$p_\| = p_{FL} \sin\theta_L = p_{FR} \sin\theta_R. \tag{6}$$

The system of equation (3) can be solved in a mixed representation [11], real-space for the variable $z$, and Fourier-transformed over the variable $\rho$. The formal solution reads:

$$f_{sL}(z<0) = -g_{aL} + \frac{1}{l_{zL}} \int_{-\infty}^{z} e^{-\kappa_L(z-\xi)} \langle f_{sL}(\xi) \rangle_{\theta_L} d\xi, \tag{7}$$

$$f_{sR}(z>0) = g_{aR} + \frac{1}{l_{zR}} \int_z^\infty e^{-\kappa_R(\xi-z)} \langle f_{sR}(\xi) \rangle_{\theta_R} d\xi, \tag{8}$$

where $f_{si}(\varepsilon) = g_{si}(\varepsilon) - 2\tanh(\varepsilon_i/2T)$, $\kappa_i = [1 - i(\mathbf{k}\mathbf{l}_{i\parallel})]/l_{zi}$. We omit the common spin label to discharge a bit complexity of notations. The above solution allows mean free paths as well as Fermi momenta to be non-equivalent in the contacting ferromagnets. To get the antisymmetric GF $g_a$ in a closed form from the solution (7) and (8) we average it over a solid angle at each half-space:

$$\langle f_{sL}(z<0) \rangle_{\theta_L} = -\langle g_{aL} \rangle_{\theta_L} + \int_{-\infty}^z \left\langle \frac{e^{-\kappa_L(z-\xi)}}{l_{zL}} \langle f_{sL}(\xi) \rangle_{\theta_L} \right\rangle_{\theta_L} d\xi, \tag{9}$$

$$\langle f_{sR}(z>0) \rangle_{\theta_R} = \langle g_{aR} \rangle_{\theta_R} + \int_z^\infty \left\langle \frac{e^{-\kappa_R(\xi-z)}}{l_{zR}} \langle f_{sR}(\xi) \rangle_{\theta_R} \right\rangle_{\theta_R} d\xi. \tag{10}$$

Then, we take the slowly varying symmetric GFs $\langle f_{sL}(\xi) \rangle_{\theta_L}$ and $\langle f_{sR}(\xi) \rangle_{\theta_R}$ out of the integrals in Eqs. (9) and (10) in the point $\xi = z$. The resulting linear equations provide approximate expressions for the angular-averaged symmetric GFs:

$$\langle f_{si}(z) \rangle_{\theta_i} = sgn(z) \frac{\langle g_{ai} \rangle_{\theta_i}}{1-\lambda_i}, \tag{11}$$

where

$$\lambda_i = \int_0^\infty \left\langle \frac{e^{-\kappa_i \eta}}{l_{zi}} \right\rangle_{\theta_i} d\eta = \frac{1}{kl_i} \arctan(kl_i). \tag{12}$$

Now, solution (11) can be used in combination with Eqs. (7) and (8) to satisfy the boundary condition Eqn. (5). Consecutive substitution of Eqn. (11) into Eqs. (7) and (8), and then, the result into BC (5) gives:

$$2g_a = -2D\left[\tanh\left(\frac{\varepsilon}{2T}\right) - \tanh\left(\frac{\varepsilon-eV}{2T}\right)\right]\gamma_k$$
$$- \frac{\langle g_{aL} \rangle_{\theta_L}}{1-\lambda_L} \int_0^\infty D\frac{e^{-\kappa_L \eta}}{l_{zL}} d\eta - \frac{\langle g_{aR} \rangle_{\theta_R}}{1-\lambda_R} \int_0^\infty D\frac{e^{-\kappa_R \eta}}{l_{zR}} d\eta, \tag{13}$$

where

$$\gamma_k = \int_0^a d\rho \int_0^{2\pi} \rho e^{i\mathbf{k}\rho} d\varphi = \frac{2\pi a}{k} J_1(ka). \tag{14}$$

Assuming first the antisymmetric GF $g_a$ in the left-hand side of (13) being equal to $g_{a1}$, according to BC (4), and then to $g_{a2}$, after the solid-angle averaging in a proper half-space we arrive at a system of two equations the solution of which looks a bit cumbersome:

$$\langle g_{aL} \rangle_{\theta_L} = -2\left[\tanh\left(\frac{\varepsilon}{2T}\right) - \tanh\left(\frac{\varepsilon-eV}{2T}\right)\right]\gamma_k$$
$$\times \left\{ \langle D \rangle_{\theta_L} [2(1-\lambda_L)(1-\lambda_R) + \tilde{\lambda}_2(1-\lambda_L)] - \langle D \rangle_{\theta_R} \tilde{\lambda}_4(1-\lambda_L) \right\} Den^{-1}(k), \tag{15}$$

$$\langle g_{aR} \rangle_{\theta_R} = -2\left[\tanh\left(\frac{\varepsilon}{2T}\right) - \tanh\left(\frac{\varepsilon-eV}{2T}\right)\right]\gamma_k$$

$$\times \left\{ \langle D \rangle_{\theta_R} [2(1-\lambda_L)(1-\lambda_R) + \tilde{\lambda}_1(1-\lambda_R)] - \langle D \rangle_{\theta_L} \tilde{\lambda}_3(1-\lambda_R) \right\} Den^{-1}(k), \qquad (16)$$

where

$$Den(k) = 4(1-\lambda_L)(1-\lambda_R) + 2\left[\tilde{\lambda}_1(1-\lambda_R) + \tilde{\lambda}_2(1-\lambda_L)\right] - \tilde{\lambda}_3\tilde{\lambda}_4 + \tilde{\lambda}_1\tilde{\lambda}_2, \qquad (17)$$

$$\tilde{\lambda}_1 = \int_0^\infty \left\langle D \frac{e^{-\kappa_L \eta}}{l_{zL}} \right\rangle_{\theta_L} d\eta \equiv \langle I_L \rangle_{\theta_L}, \qquad (18)$$

$$\tilde{\lambda}_2 = \langle I_R \rangle_{\theta_R}, \quad \tilde{\lambda}_3 = \langle I_L \rangle_{\theta_R}, \quad \tilde{\lambda}_4 = \langle I_R \rangle_{\theta_L}. \qquad (19)$$

Being substituted into the right-hand side of equation (13), equations (15) and (16) solve the problem of finding the current density, Eqn. (2), and eventually the net current $I^z$ (1) through the nanocontact as follows (a linear approximation on the applied bias voltage $V$ has been utilized):

$$I^z(z \to 0, t = 0) = \frac{e^2 p_{FL}^2 \pi a^2 V}{\pi^2} \int_0^\infty dk \frac{J_1^2(ka)}{k} F(k), \qquad (20)$$

where

$$F(k) = \langle x_L D \rangle_{\theta_L} - \left( G_1 \langle x_L I_L \rangle_{\theta_L} + G_2 \langle x_L I_R \rangle_{\theta_L} \right), \qquad (21)$$

$$G_1 = \left\{ \langle D \rangle_{\theta_R} [2(1-\lambda_R) + \tilde{\lambda}_2] - \langle D \rangle_{\theta_L} \tilde{\lambda}_4 \right\} Den^{-1}(k), \qquad (22)$$

$$G_2 = \left\{ \langle D \rangle_{\theta_R} [2(1-\lambda_L) + \tilde{\lambda}_1] - \langle D \rangle_{\theta_L} \tilde{\lambda}_3 \right\} Den^{-1}(k), \qquad (23)$$

$$\langle x_L I_L \rangle_{\theta_L} = \int_0^\infty \left\langle \cos\theta_L D \frac{e^{-\kappa_L \eta}}{l_{zL}} \right\rangle_{\theta_L} d\eta, \langle x_L I_R \rangle_{\theta_L} = \int_0^\infty \left\langle \cos\theta_L D \frac{e^{-\kappa_R \eta}}{l_{zR}} \right\rangle_{\theta_L} d\eta. \qquad (24)$$

Notice finally that the current given by equation (20) refers to a particular spin-channel of conductance in spite of the spin index is omitted for brevity. The total current through the nanocontact is the sum of currents for the both spin-channels. Formal expression for the second one is the same, but with all physical parameters referred to the second spin-channel (see the next Section).

## 3. Magnetoresistance of ferromagnetic nanocontacts

### 3.1. General considerations

The total current through a magnetic nanocontact combines two spin-channels whose conductances are different for P and AP mutual orientations of magnetisations in the banks. The magnetoresistance is characterized by a dimensionless ratio:

$$MR = \frac{\sigma^P - \sigma^{AP}}{\sigma^{AP}} \times 100\%, \qquad (25)$$

where $\sigma^{P(AP)} = \sigma_\uparrow^{P(AP)} + \sigma_\downarrow^{P(AP)} (I^{P(AP)} = \sigma^{P(AP)} V)$. Then, MR is positive if the physical effect itself is negative (the resistance drops when magnetic field is applied). Now the dependence of MR on the conduction band parameters of contacting ferromagnets can be analysed.

To account for a finite nanocontact length, we place linear-profile DW inside the nanocontact for the AP alignment of magnetisations [12,18-20]. The quantum-mechanical transmission coefficient through the linear DW can be expressed as follows:

$$D^{SL}(x_L, L) = \frac{4 k_M k_m t^2(L) \pi^{-2}}{(k_M \beta - k_m \gamma)^2 + (k_M k_m \alpha + \kappa)^2}, \quad (26)$$

where

$$\begin{aligned}
\alpha &= Ai(q_1 L) Bi(q_2 L) - Bi(q_1 L) Ai(q_2 L), \\
\beta &= t(L)\{Ai(q_1 L) Bi'(q_2 L) - Bi(q_1 L) Ai'(-q_2 L)\}, \\
\gamma &= t(L)\{Ai'(q_1 L) Bi(q_2 L) - Bi'(q_1 L) Ai(q_2 L)\}, \\
\kappa &= t^2(L)\{Ai'(q_1 L) Bi'(q_2 L) - Bi'(q_1 L) Ai'(q_2 L)\},
\end{aligned} \quad (27)$$

and $t(L) = [2mE_{ex}/L]^{1/3}$, $E_{ex} = (k_{FM}^2 - k_{Fm}^2)/2m$, $q_1 = -k_{FM}^2 t(L)/2mE_{ex}$, $q_2 = -k_{Fm}^2 t(L)/2mE_{ex}$, where $L$ is the width of DW; $Ai(z)$, $Bi(z)$, $Ai'(z)$, and $Bi'(z)$ are the Airy functions and their derivatives; $k_m = k_{Fm} \cos(\theta_m)$ and $k_M = k_{FM} \cos(\theta_M)$ are the normal components of the wave vector of minority and majority subband, respectively. Note here that $k_m$ is used for a subband with smaller Fermi momentum, and $k_M$ for a subband with larger Fermi momentum whatever the spin projection of the subband, or the side of the contact - left or right, is. The quantum-mechanics textbook expression for the coefficient of transmission through a step-like DW (band-offset model), $D^{step}(x_L) = 4 k_M k_m/(k_M + k_m)^2$, can be retrieved from equation (26) in the limit of $L \to 0$. Again, we omit the spin index to simplify appearance of the formulas above.

Ferromagnetic heterocontacts mean that the contacting ferromagnets have different parameters of their conduction bands. In our calculations we fix parameters of the ferromagnetic metal at the left bank of the contact (the values $k_{F\downarrow} = 4$ nm$^{-1}$, $k_{F\uparrow} = 10$ nm$^{-1}$, $k_{F\downarrow}/k_{F\uparrow} = 0.4$, are close to that for iron cited in references [21,22]), and vary the conduction band properties of the second ferromagnetic metal. Before we proceed with particular calculations, we have to mention an important detail which distinguishes the ferromagnetic heterocontacts from homocontacts. At parallel configuration of magnetisations in a homocontact there is no DW in the constricted area, and electron of either spin-projection moves in a flat potential landscape because materials (conduction bands) on the both sides of the contact are identical. Then, the quantum-mechanical transmission coefficient $D$ in both conduction spin-channels is equal to one. In contrast in a ferromagnetic heterocontact there is always a potential barrier at nanocontact because the conduction band bottoms do not coincide for either spin projection and magnetisation alignments. The only difference is that in the P magnetisation configuration there is no DW in the nanocontact, but in the AP configuration there is. Then, for the P alignment we have to assume sharp change in the band parameters just at the interface of two ferromagnets, but for the AP alignment we place linear profile DW inside the nanocontact. In a common stream of numerical calculation we simply simulate the sharp interface in the P configuration by a linear DW of about one angström of thickness ($L = 0.1$ nm).

## 3.2. Magnetoresistance of a ferromagnetic heterocontact

In a simple parabolic band we use here the heterocontacts are sorted by mutual positions of bottoms of their conduction spin-subbands at the parallel alignment of magnetisations. Three physically distinct combinations can be considered (see insets in the figures below), and MR for every combination is calculated as a function of the contact size. Looking through the insets of the figures one may see that the Fermi momentum of a spin-subband from the left to the contact can be larger as well as smaller than that from the right, however, according to the momentum conservation law Eq. (6), not every incident angle from the side of the bigger momentum is allowed for electron to transmit to the side of the smaller Fermi momentum. Then, the integrals in Eqs. (18), (19) and (24) can be evaluated as follows (the momentum conservation law (6) has been used to bring the integration variables to the left-hand side angle $\theta_L$):

$$\langle D \rangle_{\theta_L} = \int_{x_c}^{1} dx_L D(x_L), \quad \langle D \rangle_{\theta_R} = \int_{x_c}^{1} dx_L \frac{x_L D(x_L) \delta}{\sqrt{x_L^2 \pm x_{cr}^2}}, \quad \langle x_L D \rangle_{\theta_L} = \int_{x_c}^{1} dx_L D(x_L) x_L, \tag{28}$$

$$\langle x_L I_L \rangle_{\theta_L} = \int_{x_c}^{1} \frac{x_L D(x_L) dx_L}{\sqrt{1 + (kl_L)^2 (1 - x_L^2)}}, \quad \langle x_L I_R \rangle_{\theta_L} = \int_{x_c}^{1} \frac{x_L D(x_L) dx_L}{\sqrt{1 + (kl_R \delta)^2 (1 - x_L^2)}}, \tag{29}$$

$$\tilde{\lambda}_1 = \int_{x_c}^{1} dx_L \frac{D(x_L)}{\sqrt{1 + (kl_L)^2 (1 - x_L^2)}}, \tag{30}$$

$$\tilde{\lambda}_2 = \int_{x_c}^{1} dx_L \frac{x_L D(x_L) \delta}{\sqrt{x_L^2 \pm x_{cr}^2} \sqrt{1 + (kl_R \delta)^2 (1 - x_L^2)}}, \tag{31}$$

$$\tilde{\lambda}_3 = \int_{x_c}^{1} dx_L \frac{x_L D(x_L) \delta}{\sqrt{x_L^2 \pm x_{cr}^2} \sqrt{1 + (kl_L)^2 (1 - x_L^2)}}, \tag{32}$$

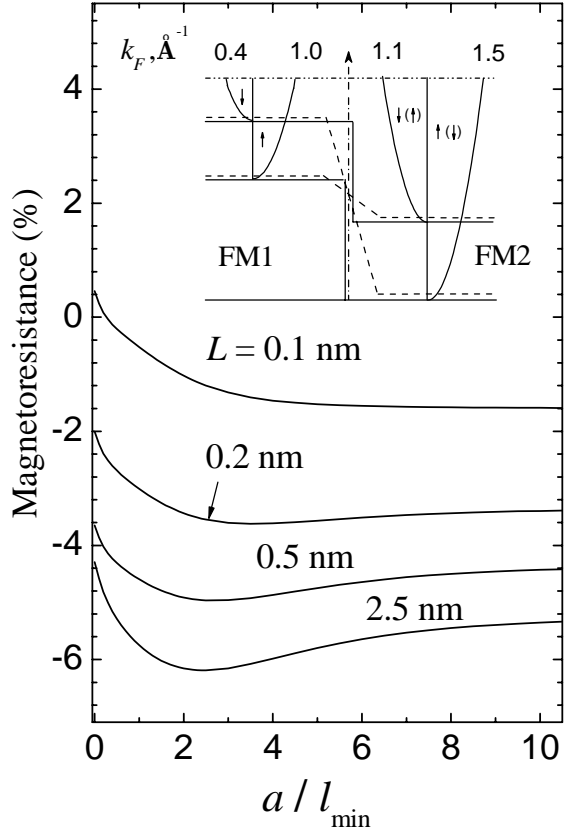

Fig. 1. Dependence of MR on the contact size for the case $k_{FL\downarrow} < k_{FL\uparrow} < k_{FR\downarrow} < k_{FR\uparrow}$. The Fermi momenta of the contacting ferromagnets are indicated in the inset, the ratio $l_{L\downarrow}/l_{L\uparrow} = 2$ is taken equal to $l_{R\downarrow}/l_{R\uparrow}$ to simplify appearance, and $l_{L\downarrow}/l_{R\downarrow} = 1$.

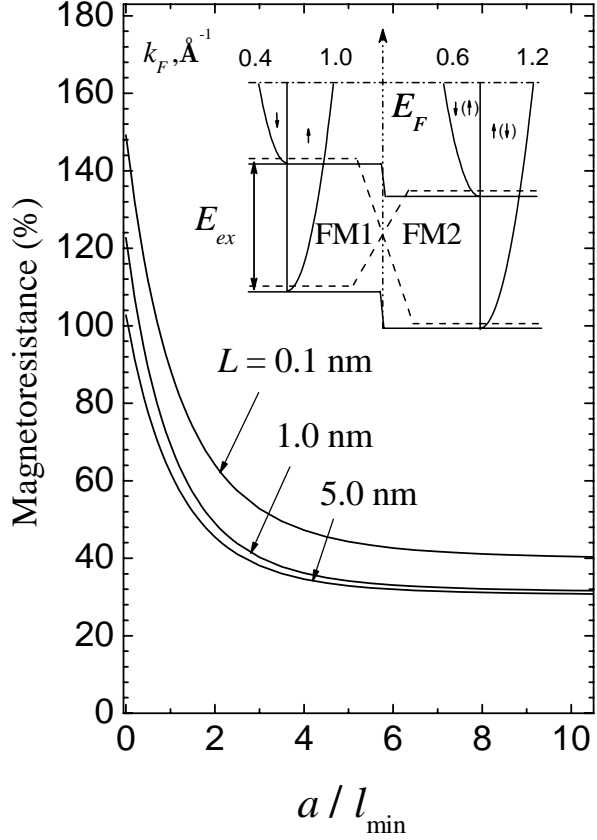

Fig. 2. Dependence of MR on the contact size for the case $k_{FL\downarrow} < k_{FR\downarrow} < k_{FL\uparrow} < k_{FR\uparrow}$. The layout and choice of the mean free paths are the same as in Fig. 1.

$$\tilde{\lambda}_4 = \int_{x_c}^{1} dx_L \frac{D(x_L)}{\sqrt{1+(kl_R\delta)^2(1-x_L^2)}}, \qquad (33)$$

where $\delta = k_{FL}/k_{FR}$. If $k_{FL} < k_{FR}$ then $x_c = 0$, $x_{cr} = \sqrt{(1-\delta^2)/\delta^2}$, and the upper sign in the square roots has to be used. When $k_{FL} > k_{FR}$, $x_c = x_{cr}$, $x_{cr} = \sqrt{1-(\delta)^{-2}}$, and the lower sign in the square roots has to be used.

Figure 1 displays MR of a hypothetical heterocontact in which the right-hand side ferromagnet has larger Fermi momenta for both conduction spin-subbands compared with those for the left-hand side one ($k_{FL\downarrow} < k_{FL\uparrow} < k_{FR\downarrow} < k_{FR\uparrow}$). In contrast to the case of homocontact, MR of this type of heterocontact is negative, and decreases in the absolute value when approaching the ballistic regime ($a < l_{min}$, where $l_{min} = l_{L\uparrow}$, and also for figures 2 and 3). Moreover, there is a shallow valley in the range of $a/l_{min} \approx 1-6$. Positive physical magnetoresistance (to which the negative MR values are given in figure 1 because of the definition, equation (25)) is explained by more sharp potential barriers (more resistive interface) in the P alignment compared with the smoothen potential landscape in the presence of a domain wall (see the inset in figure 1). The magnitude of the magnetoresistance effect is rather small.

Magnetoresistance of the second type of heterocontact ($k_{FL\downarrow} < k_{FR\downarrow} < k_{FL\uparrow} < k_{FR\uparrow}$) is displayed in figure 2. MR is positive in the entire range of the contact sizes and much bigger in magnitude compared with the first case given in Fig. 1. It increases about four times upon changing the conductance regime on the size of the nanocontact from diffusive to ballistic.

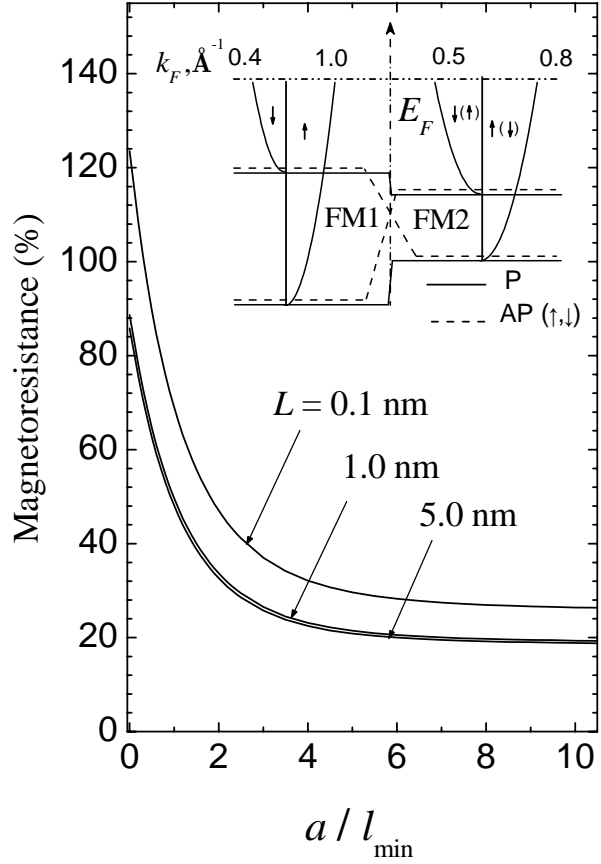

Fig. 3. Dependence of MR in the contact size for the case $k_{FL\downarrow} < k_{FR\downarrow} < k_{FR\uparrow} < k_{FL\uparrow}$. The layout and choice of the mean free paths are the same as in Fig. 1.

Figure 3 shows dependence of MR on the contact size for case 3 ($k_{FL\downarrow} < k_{FR\downarrow} < k_{FR\uparrow} < k_{FL\uparrow}$). The MR behaviour is similar to the second case, the sign of MR is always positive (magnetoresistance is negative). Considerable enhancement of MR follows from the calculations upon approaching the ballistic regime of conductance in vicinity of the nanocontact. It is worth noting here that if we exchange spin indices of all spin-dependent quantities in the formulas above the MR($a$) dependences in figures 1-3 do not change.

## 3.3. Discussion of experiments

To the authors' knowledge there are three reports on magnetoresistance measurements in ferromagnetic contacts: Mumetal-Ni [23,4] and $CrO_2$-Ni [24]. Mumetal ($Ni_{77}Fe_{14}Cu_5Mo_4$) is close to Permalloy ($Ni_{79}Fe_{21}$) on its composition. Therefore, we may use the material parameters of permalloy and nickel [16,25-31] as a trial guess to calculate MR of the Mumetal-Ni couple. The results for MR of Mumetal-Ni couple correspond to the case 3 are given in the figure 4. The ballistic limit magnitude of MR varies in the range 75-100% ($L = 0.1$-5 nm) with parameters which are given in figure 4 and 82-132% with next values $k_{FL\uparrow} = 0.56$, $k_{FL\downarrow} = 1.1$, and $k_{FR\uparrow} = 0.6$, $k_{FR\downarrow} = 1.08$ Å$^{-1}$. It is agree satisfactorily with the experimental values of MR = 78-132% quoted in Table 1 of Refs. [23,4] and Fig. 2 in Ref. [4] at smallest conductances for the P-alignment of magnetisations. As for the case of $CrO_2$-Ni heterocontact [24], we would abstain from considering the data in the frame of the present calculations, because the parallel alignment conductance is too low to treat the Ni-$CrO_2$ nanocontacts as being true metallic conducting ones. Tunnelling conductance regime, which we suspect in Ni-$CrO_2$ heterocontacts, is beyond the scope of our theory.

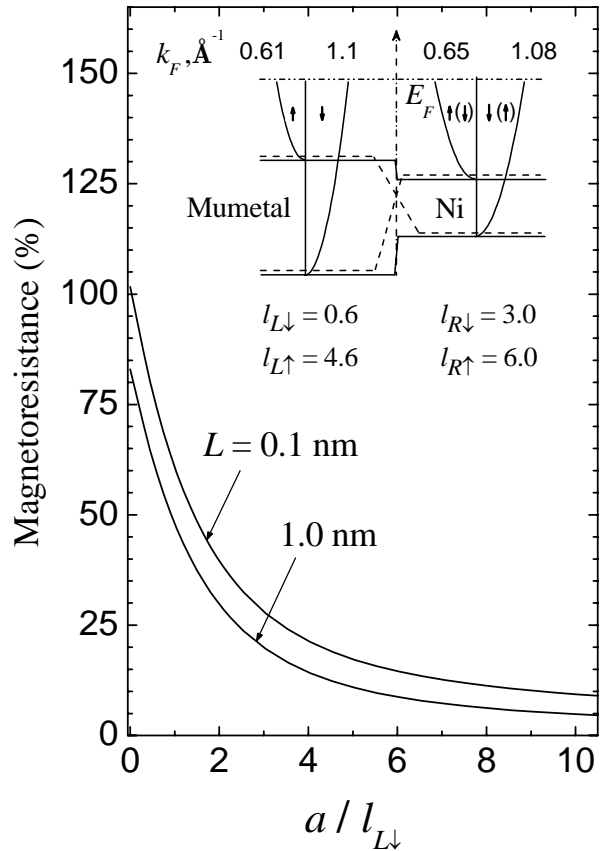

Fig. 4. Dependence of MR on the contact size for the choice of parameters close to the Mumetal-Ni heterocontact. The values of parameters are given in the figure.

## 4. Conclusions

To summarize, in this paper we investigated theoretically GMR in nanoscale ferromagnetic heterocontacts. The quasiclassical theory of magnetic nanocontacts was generalized for the case of metallic ferromagnets with arbitrary Fermi momenta and mean free paths of the conduction spin-subbands. The heterocontacts were sorted by three types of mutual positions of conduction spin-subband bottoms. A model of linear domain wall profile for antiparallel alignment of magnetisations in contacting ferromagnets was used to account for the finite contact length. In general, the magnetoresistance plotted against size of the nanocontact can be of either sign depending on the conduction bands matching. The magnitude of the effect for heterocontacts in our calculations was always smaller than that for a contact made of the same ferromagnetic metal. The magnetoresistance in the case, when one of the ferromagnetic metals has both Fermi momenta of the conduction electron spin-subbands smaller than the other, is always much smaller compared with the other band arrangements considered in the paper. The theoretical results agree satisfactorily with the available experimental data on the ferromagnetic heterocontacts Mumetal-Ni.
The work was supported by the EC grant NMP4-CT-2003-505282.